\documentclass[aps,floatfix,twocolumn,showpacs,pra]{revtex4}
\usepackage{graphicx,bm}
\usepackage{amsmath}

\begin{document}
\title{Genuine phase diffusion of a Bose-Einstein condensate in the
microcanonical ensemble: A classical field study}
\author{A. Sinatra}
\affiliation{Laboratoire Kastler Brossel,
Ecole Normale Sup\'erieure, UPMC and CNRS,
24 rue Lhomond, 75231 Paris Cedex 05, France}
\author{Y. Castin}
\affiliation{Laboratoire Kastler Brossel, 
Ecole Normale Sup\'erieure, UPMC and CNRS,
24 rue Lhomond, 75231 Paris Cedex 05, France}

\begin{abstract}
Within the classical field model, we find that the phase of a Bose-Einstein condensate undergoes a true diffusive motion
in the microcanonical ensemble, the variance of the condensate phase change between time zero and time $t$ growing linearly in $t$.
The phase diffusion coefficient obeys a simple scaling law in the double thermodynamic and Bogoliubov limit.
We construct an approximate calculation of the diffusion coefficient, in fair agreement with the numerical results
over the considered temperature range, and we extend this approximate calculation to the quantum field.
\end{abstract}


\maketitle
\section{Introduction}
Phase coherence is one of the most prominent properties of Bose-Einstein condensates, relevant for applications
of condensates in metrology and quantum information \cite{metrology_quantuminfo}. 
The issue of condensate phase dynamics and phase spreading at zero temperature due to interactions has been extensively studied
in theory \cite{PhaseT=0} and experiments \cite{MIT,JILA,other_phase_expts}. There is a renewed interest in this issue of temporal 
phase coherence due to the recent studies in low dimensional quasi condensates, both experimentally \cite{Ketterle1D,Widera,Schmidtmayer} 
and theoretically \cite{Demler}. The present work addresses the problem of 
the determination of fundamental limits of phase coherence in a true three dimensional Bose-Einstein condensates at non zero temperature.

The effect of finite temperature on phase coherence in a Josephson junction realized by a condensate trapped in a double well
potential has been studied in \cite{Pitaevskii} and \cite{Oberthaler}. 
The situation is different when the two condensates are separated. In
this case there is no restoring force for the relative phase which then evolves independently in the two BECs  \cite{Sols}. 
The effect of the temperature in this case, joint to the effect of the interactions, is to provide a spreading in time of the relative phase.  

In a previous work \cite{PRAEmilia}, we considered a condensate prepared in an equilibrium state in the canonical ensemble.
In that case we could show using ergodicity that the phase change of the condensate during a time $t$ has a variance which grows proportionally to $t^2$. 
In other words, the condensate phase spreading in the canonical ensemble is ballistic  \cite{Kuklov} and not diffusive  \cite{Zoller,GrahamPRL,GrahamPRA,GrahamJMO}. 
As we could calculate in \cite{PRAEmilia} using an ergodic theory, the coefficient of this super-diffusive thermal spreading
is proportional to the variance of the energy in the considered equilibrium state. 
If we now suppress the fluctuations of energy in the initial state, by moving
from the canonical ensemble to the microcanonical ensemble, the ballistic thermal spreading disappears and one may expect that the condensate
phase undergoes a genuine diffusion in time. 
In the present work we show that this is indeed the case and we study this genuine phase diffusion numerically within the classical
field model \cite{Kagan,Sachdev,Drummond,Rzazewski0,Burnett,Wigner} described in section \ref{sec:model}. 
The numerical results are presented in section \ref{sec:num}, and their analysis shows the existence of simple
scaling laws and of an universal curve giving the phase diffusion coefficient in the double thermodynamical limit ($N\to \infty$, 
volume $V\to\infty$
density $\rho=\mbox{constant}$) and Bogoliubov
limit ($N\to \infty$, coupling constant $g\to 0$, $Ng=\mbox{constant}$). In section \ref{sec:heuristic}
 we derive an  aproximate formula for the diffusion coefficient, that we compare to the numerical results 
 and  that we also extend to the quantum field. We conclude in section \ref{sec:conclusion}.

\section{Classical field model and numerical procedure}
\label{sec:model}
We consider a lattice model for a classical field $\psi({\bf r})$ in
three dimensions. The lattice spacings are $l_1$, $l_2$, $l_3$ 
along the three directions 
of space and $dV=l_1 l_2 l_3$ is the volume of the unit cell in the lattice.
We enclose the atomic field in a spatial box of sizes $L_1$, $L_2$, $L_3$
and volume $V=L_1 L_2 L_3$, with periodic boundary conditions.
To guarantee efficient ergodicity in the system we choose non 
commensurable square lengths in the ratio
$L_1^2 : L_2^2 : L_3^2 =  \sqrt{2} : (1+\sqrt{5})/2 : \sqrt{3}$. 
The lattice spacings squared $l_1^2$, $l_2^2$, $l_3^2$ are in the same ratio. 

The field $\psi$ may be expanded over the plane waves
\begin{equation}
\psi({\bf r})=\sum_{\bf k} a_{\bf k} 
\frac{e^{ \, i \, {\bf k}\cdot{\bf r}}}{\sqrt{V}} \,,
\label{eq:planewaves}
\end{equation}
where $\mathbf{k}$ is restricted to the first Brillouin zone,
$k_\alpha\in [-\pi/l_\alpha,\pi/l_\alpha[$ and $\alpha$ labels the directions
of space.

We assume that, in the real physical system, the total number of atoms is fixed, equal to $N$.
In the classical field model, this fixes the norm squared of the field:
\begin{equation}
dV \sum_{\mathbf{r}} |\psi(\mathbf{r})|^2 = N.
\end{equation}
Equivalently the density of the system 
\begin{equation}
\rho=\frac{N}{V}
\end{equation}
is fixed for each realization of the field.
The evolution of the field is governed by the Hamiltonian
\begin{equation}
H=\sum_{\bf k} \tilde{E}_k {a}_{\bf k}^* a_{\bf k}
+ \frac{g}{2} \sum_{{\bf r}} dV
\psi^*({\bf r})\psi^*({\bf r})
\psi({\bf r}) \psi({\bf r}),
\label{eq:Hamiltonian}
\end{equation}
where $\tilde{E}_k$ is the dispersion relation of the non-interacting waves,
and the binary interaction between particles in the real gas is reflected in the classical
field model by a field self-interaction with a coupling constant
\begin{equation}
g=\frac{4\pi \hbar^2 a}{m}\,, 
\end{equation}
where $a$ is the $s$-wave scattering length of
two atoms.

As a matter of a fact we use here the same refinement as in \cite{PRAEmilia} consisting in modifying the dispersion relation in order to obtain
for the ideal gas the correct quantum values of the mean occupation numbers at equipartition
\begin{equation}
\frac{1}{e^{\beta \hbar^2 k^2/2m} -1} = \frac{k_B T}{\tilde{E}_k} \,.
\label{eq:Ektilde}
\end{equation}
However we do not expect this to have a large impact here as we put a cut-off at an energy of the order of $k_BT$.
More precisely we choose the number of the lattice points in a temperature dependent way,
such that the maximal Bogoliubov energy on the lattice is equal to $k_BT$:
\begin{equation}
{\mathrm{max}}_{\,\bf k} \: [(\hbar^2 k^2/2m)(2\rho g+\hbar^2k^2/2m)]^{1/2}=k_BT \,.
\label{eq:cutoff}
\end{equation}

The discretized field has the following Poisson brackets
\begin{equation}
i\hbar \{\psi({\bf r_1}),\psi^*({\bf r_2})\}=
	\frac{\delta_{{\bf r_1},{\bf r_2}}}{dV} 
\end{equation}
where the Poisson brackets are such that $df/dt = \{f,H\}$ for a
time-independent functional $f$ of the field $\psi$.
The field then evolves according to the non linear equation \cite{Delta}
\begin{equation}
i \hbar \, \partial_t \psi= \left\{k_B T
\left[\exp\left(-\beta\frac{\hbar^2}{2m}\Delta\right)-1
\right]
+ g |\psi({\bf r},t)|^2  \right\} \, \psi\, .
\label{eq:sc}
\end{equation}

We introduce the density and the phase of the condensate mode
\begin{equation}
a_0=e^{\, i \, \theta} \sqrt{N_0}\,.
\label{eq:intphase}
\end{equation}
The quantity of interest is the variance of the condensate phase change during $t$:
\begin{equation}
\mbox{Var}\, \varphi(t)=\langle {{\varphi}}(t)^2 \, \rangle -
\langle {{\varphi}}(t) \, \rangle^2
\end{equation}
where
\begin{equation}
{{\varphi}}(t)={\theta}(t)-{\theta}(0).
\label{eq:varphase}
\end{equation}
The averages are taken over stochastic realizations
of the classical field, as the 
initial field samples the microcanonical ensemble with an energy $E$. For convenience, we parametrize the microcanonical ensemble
by the temperature $T$ such that the mean energy of the field in the
canonical ensemble at temperature $T$ is equal to $E$. 

To generate the stochastic initial values of the classical field
we proceed as follows. 
(i) First  we generate 1000 stochastic fields in the canonical ensemble at temperature $T$, as explained in \cite{PRAEmilia},
and we compute the
average energy of the field $\langle E \rangle_{\mathrm{can}}$ and its root mean squared fluctuations $\sigma=\sqrt{\mbox{Var} \; E}$.
(ii) We generate other fields, still in the canonical ensemble, and we filter them keeping only realizations  with an energy $E$
such that $|E-\langle E \rangle_{\mathrm{can}}|\le 0.01 \, \sigma/2$.
(iii) We let each field evolve for some time interval with the Eq.(\ref{eq:sc})
to eliminate transients due to the fact that the Bogoliubov approximation, used in the
sampling, does not produce an exactly stationary distribution.
After this `thermalization' period we start calculating the relevant observables,
as $\psi$ evolves with the same equation (\ref{eq:sc}).
In practice this equation is integrated numerically with the FFT splitting technique. The ensemble of data reported here has required a CPU time
of about two years on Intel Xeon Quad Core 3 GHz processors. 

\section{Numerical results and scaling laws}
\label{sec:num}
\begin{figure}[htb]
\centerline{\includegraphics[width=7cm,clip=]{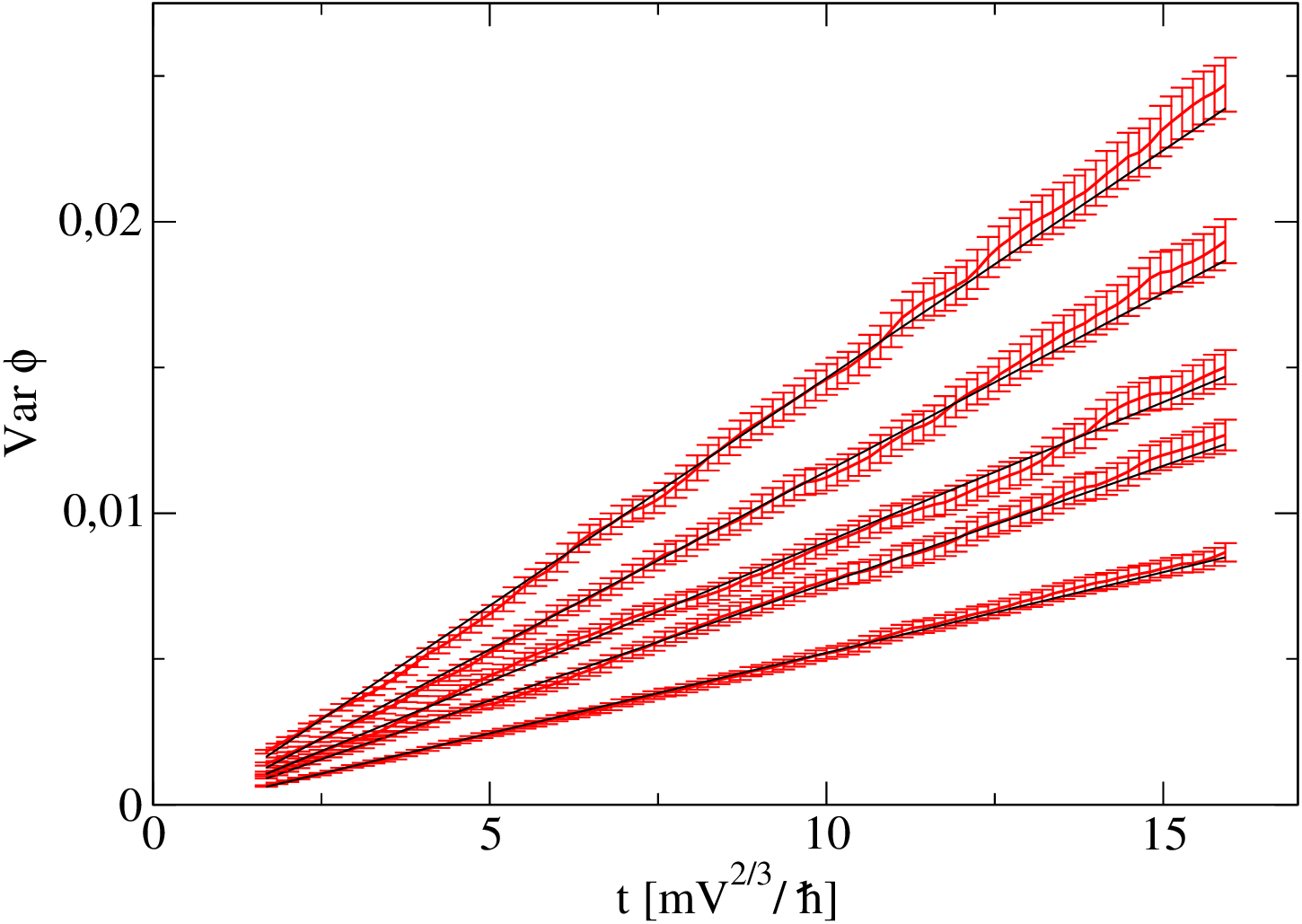}}  
\caption{Variance of the condensate phase change $\varphi(t)$
as a function of time. Wavy line with error bars: Numerical
results. Solid lines: A linear fit. 
From bottom to top, the reduced temperature $k_BT/\rho g$
is 9.7, 13.2, 15.7, 19.6, 24.2.
The number of atoms is fixed to $N=2.37\times 10^6$.
The high energy cut-off is fixed according to (\ref{eq:cutoff}), on a grid $32^3$, so that
the temperature slightly varies, from bottom to top: 
$k_B T/(\hbar^2/m V^{2/3})=16864, 16411, 16212, 16010, 15854$.
The time is in units of $mV^{2/3}/\hbar$.
\label{fig:varphi}}
\end{figure}

The first important result that we obtain is the diffusive behavior of the condensate phase.
In Fig.\ref{fig:varphi} we show an example of numerical data. From bottom to top, five values of $k_BT /\rho g$ are presented
for a constant number of atoms $N=2.36\times10^6$. The wavy line with error bars is the phase variance as a function of time obtained with about 1200 stochastic realizations  \cite{EBvarphi}. The solid line is a linear fit from which we deduce the value of the diffusion coefficient,
\begin{equation}
\mbox{Var} \, \varphi (t) \stackrel{t \to \infty}{\sim} 2 D t \,.
\end{equation}

\begin{figure}[htb]
\centerline{\includegraphics[width=7cm,clip=]{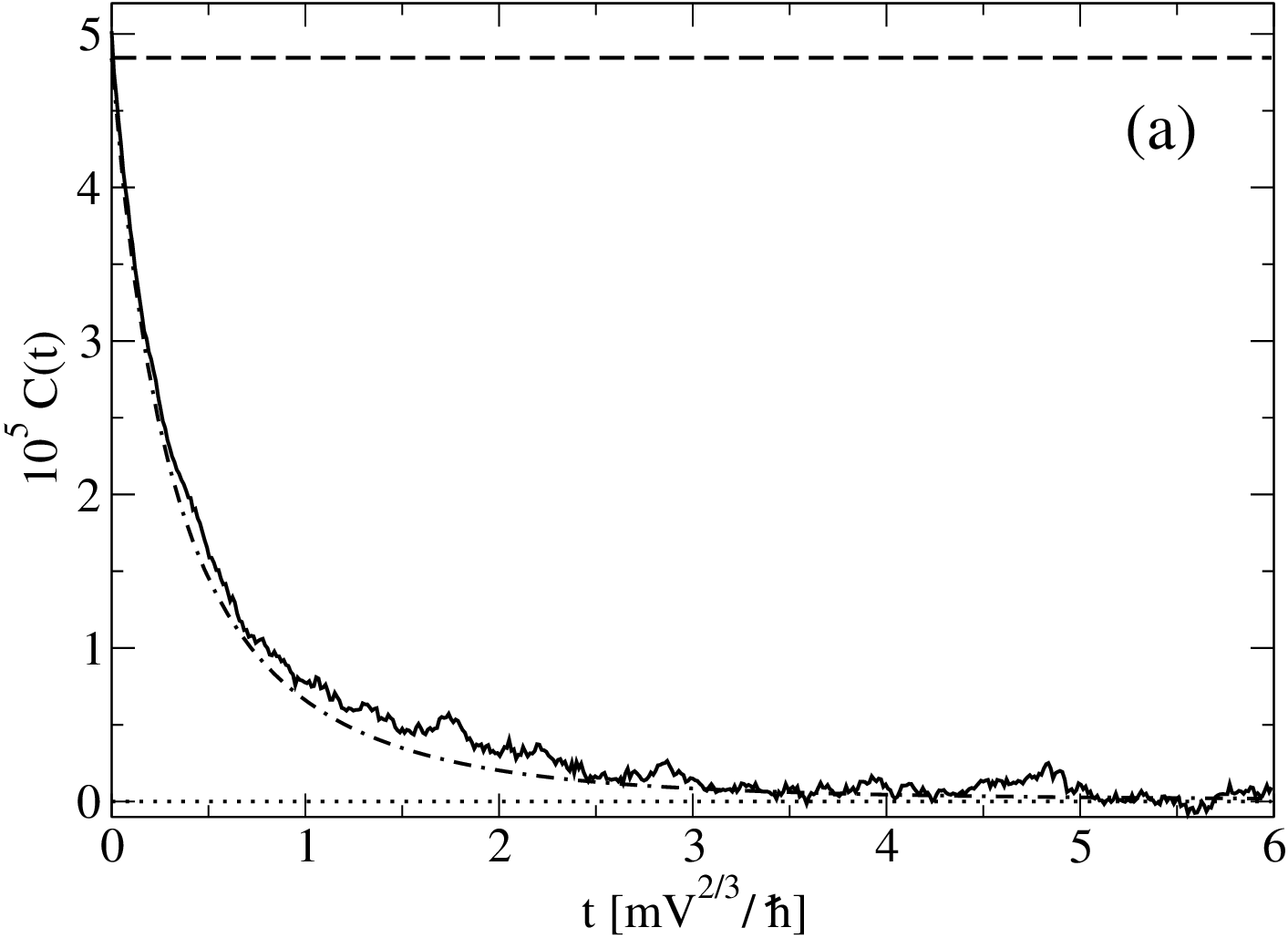}}  
\centerline{\includegraphics[width=7cm,clip=]{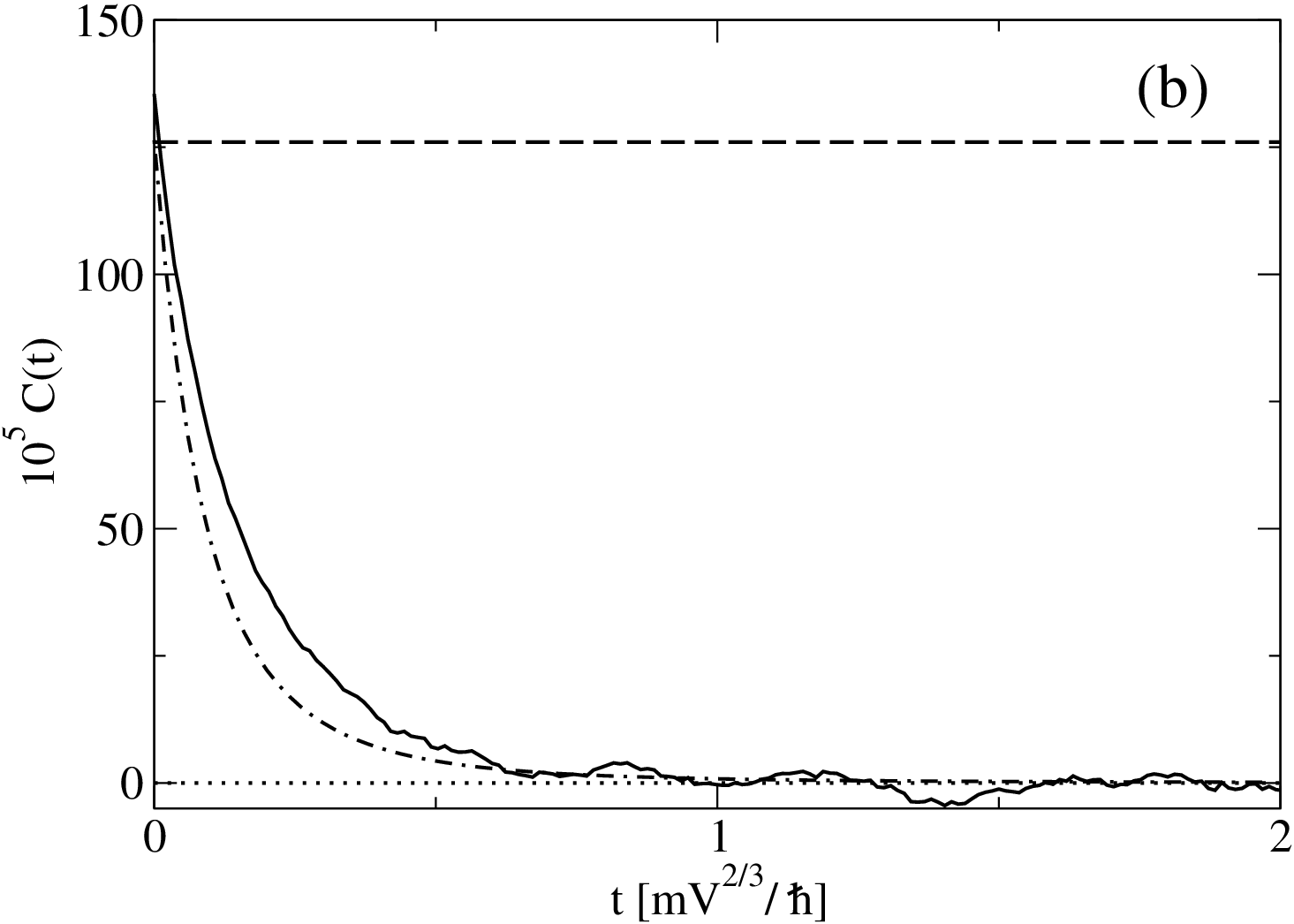}}  
\caption{Correlation function $\mathcal{C}(t)$ of the phase
derivative $\dot\varphi(t)$ given by the non-oscillating
approximation (\ref{eq:phidotnonosc}), as a function of time.
Solid line: Numerical results. Dashed line: Result of
Bogoliubov theory. Dashed-dotted line: Prediction of the projected
Gaussian approach of section \ref{sec:heuristic}.
The number of atoms is fixed to $N=5\times 10^6$, and the Gross-Pitaveskii
chemical potential is fixed to $\rho g=700 \hbar^2/m V^{2/3}$.
In (a) the temperature is $k_B T/( \hbar^2/m V^{2/3})= 5469$,
with a grid size $18^3$. 
In (b) the temperature is $k_B T/( \hbar^2/m V^{2/3})= 14054$,
with a grid size $30^3$.
\label{fig:corr}}
\end{figure}

The diffusive behavior of the condensate phase is strictly related to the long time behavior of the time correlation function $\mathcal{C}$
of the condensate phase derivative $\dot{\varphi}$, 
\begin{equation}
{\mathcal{C}}(|t^\prime-t^{\prime \prime}|)  =
\langle \dot{\varphi}(t^\prime) \dot{\varphi}(t^{\prime \prime}) \rangle -
       \langle \dot{\varphi}(t^\prime) \rangle 
       \langle \dot{\varphi}(t^{\prime \prime}) \rangle \,,
\end{equation}
where we used the fact that $\mathcal{C}$ depends only on $|t^{\prime}-t^{\prime \prime}|$ for a steady state
classical field.
By writing $\varphi(t)$ in terms of its time derivative, one obtains \cite{PRAEmilia}
\begin{equation}
\mbox{Var}\, {\varphi}(t)
 =   2t \int_0^t d \tau  \; \mathcal{C}(\tau) -2 \int_0^t d \tau  \; \tau \mathcal{C}(\tau)  \,.
\end{equation}
If  $\mathcal{C}(t)$ has a non-zero limit at long times, as it was the case in the canonical ensemble 
\cite{PRAEmilia}, $\mbox{Var}\, {\varphi}$ grows quadratically in time.
Here, in the microcanonical ensemble $\mbox{Var}\,{\varphi}$ grows linearly in time and we expect that
$\mathcal{C}(t)\rightarrow 0$ when $t \rightarrow \infty$.  
An illustration of that, for two values of the temperature, is given in Fig.\ref{fig:corr} where, for convenience,
$\mathcal{C}(t)$ is calculated with a simplified formula for the phase derivative \cite{osc_terms}
\begin{equation}
\hbar\dot{\varphi} \simeq -\rho g -
\frac{g}{V} \sum_{\mathbf{k}\neq\mathbf{0}}   \, (\tilde{U}_k+\tilde{V}_k)^2 \, |b_{\bf k}|^2 \,.
\label{eq:phidotnonosc}
\end{equation}
In (\ref{eq:phidotnonosc}) the $b_\mathbf{k}$ are the field amplitudes on the Bogoliubov modes
\cite{bk}.

\begin{figure}[htb]
\centerline{\includegraphics[width=7cm,clip=]{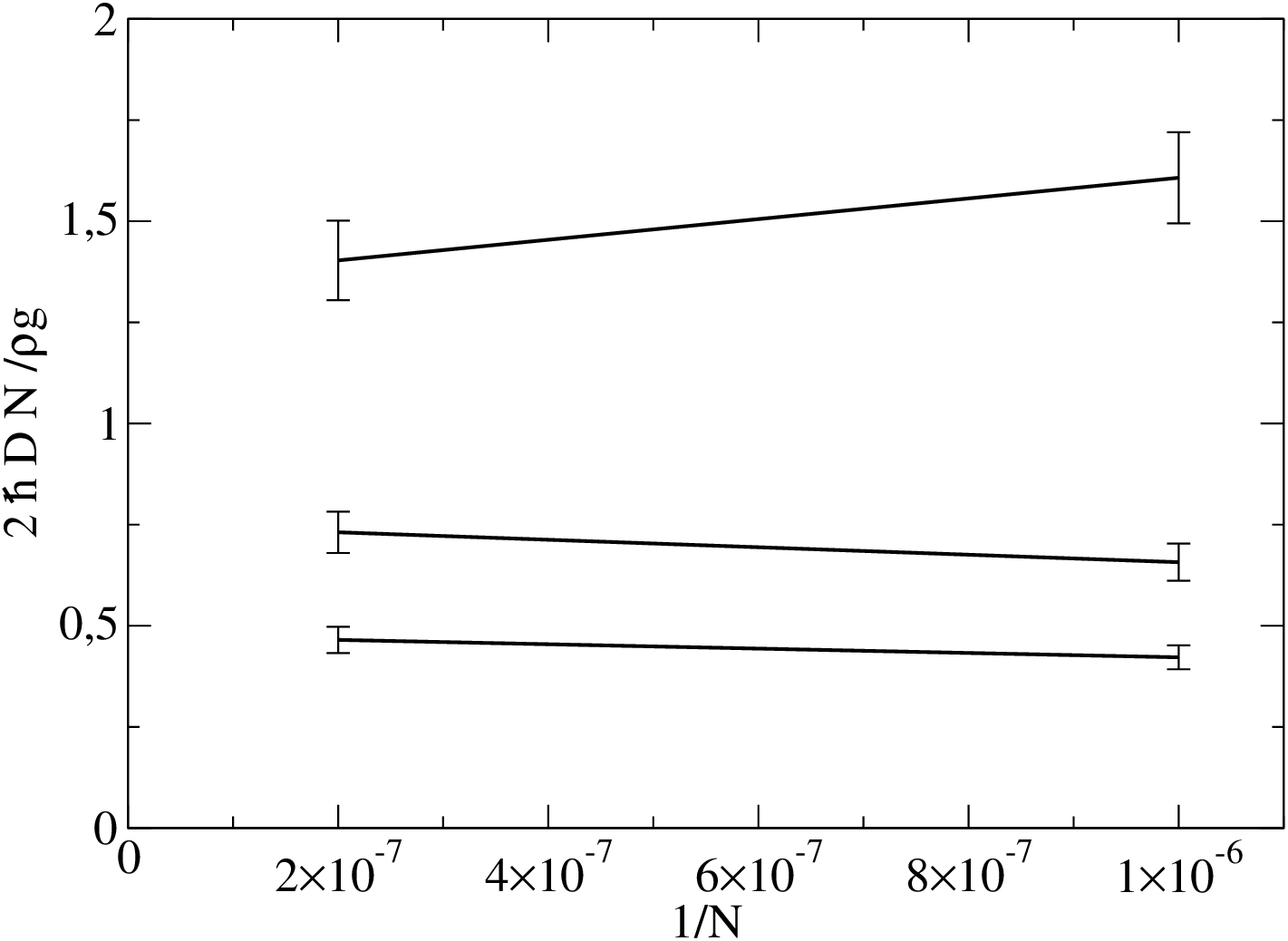}}  
\caption{
Scaling of the phase diffusion coefficient in the Bogoliubov limit
(\ref{eq:CD}), for a factor 5 variation of the atom number $N$.
The Gross-Pitaevskii chemical potential is fixed to $\rho g = 700
\hbar^2/m V^{2/3}$. Points with error bars: Simulation results.
The lines connect the points with the same temperature.
From bottom to top: $k_BT= 5469 \hbar^2/m V^{2/3}$ with a grid size
$18^3$, 
$k_BT= 6606 \hbar^2/m V^{2/3}$ with a grid size
$20^3$,
$k_BT= 9231 \hbar^2/m V^{2/3}$ with a grid size
$24^3$.
\label{fig:finale_CD}}
\end{figure}

We now investigate numerically how the diffusion coefficient scales in different limits. First we consider
the ``Bogoliubov limit" introduced in \cite{CastinDum}
\begin{equation}
N \to \infty \,, \:\:\: g \to 0 \, \:\:\:{\mbox{with}} \:\:\: Ng = \mbox{constant} \,,
\label{eq:CD}
\end{equation}
the other parameters ($V,l_1,l_2,l_3,T$) being fixed \cite{proviso}. 
In this limit, the number of non condensed
particles converges to a non zero value while the non condensed fraction vanishes. 
The time evolution of the Bogoliubov occupation numbers $n_{\mathbf{k}}=|b_{\mathbf{k}}|^2$ 
is then mainly due to terms in the interaction Hamiltonian which are  cubic in the non condensed field amplitude and linear in the
condensate amplitude and thus of order $\epsilon=g \sqrt{N}$. Physically these cubic terms describe interactions among Bogoliubov
modes such as Landau and Beliaev processes \cite{Vincent,Stringari_Pitaevskii,Giorgini,Shlyapnikov}, which are included
in the classical field model \cite{Kagan,Sachdev,DaliShlyap,Rzazewski0,Burnett,Cartago,Rzazewski1,Rzazewski2,
vortex_formation,bec_collision}.
They lead to evolution rates of the $n_{\mathbf{k}}$ of order $\epsilon^2$.
We thus expect a phase diffusion coefficient of the same order  $\epsilon^2$, which is $\propto 1/N$ according to (\ref{eq:CD}).
This expectation is confirmed numerically as we show in Fig.\ref{fig:finale_CD}, where we find that $DN$ is constant within the error bars
over a factor 5 variation of $N$ and for three considered temperatures.

\begin{figure}[htb]
\centerline{\includegraphics[width=7cm,clip=]{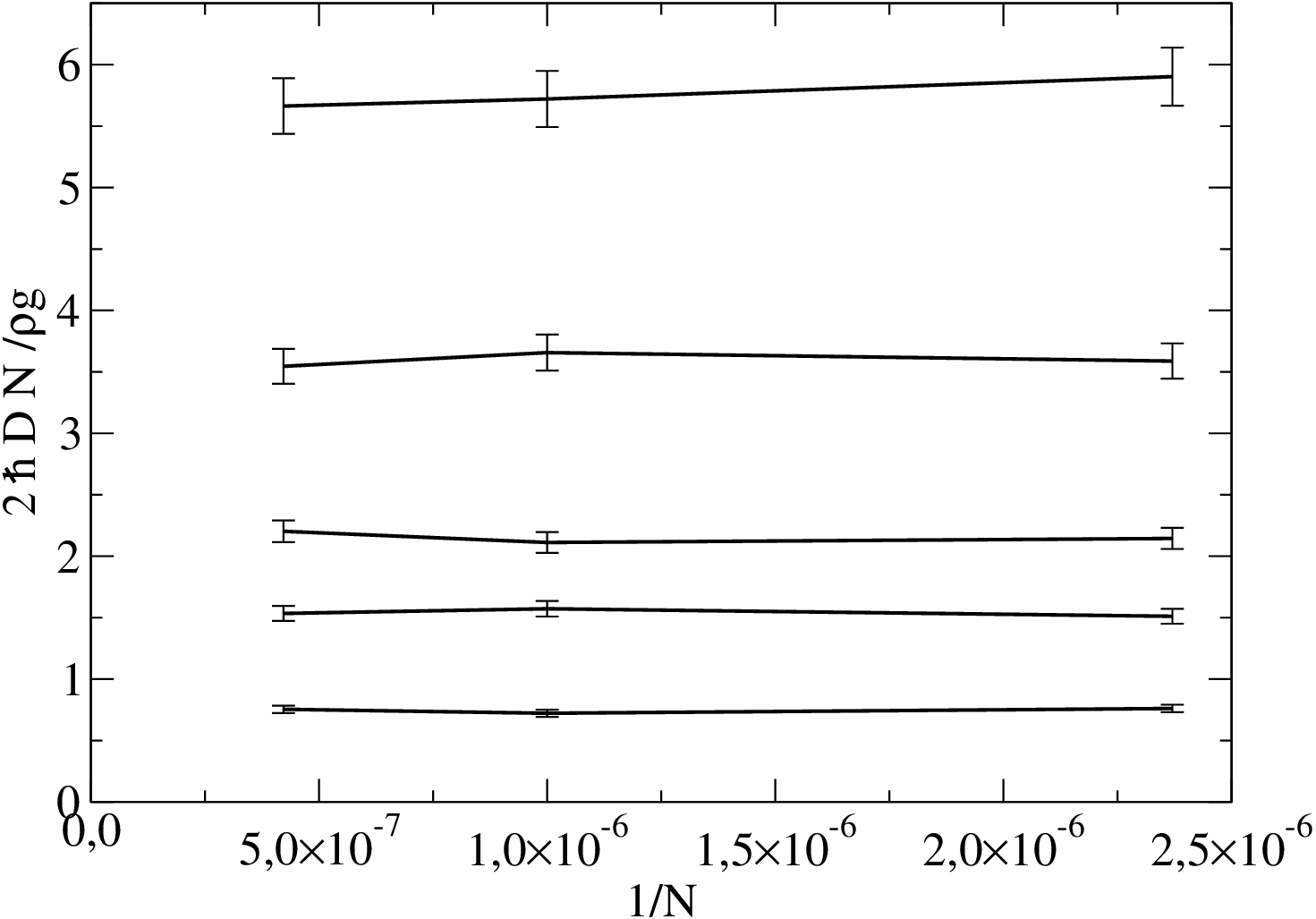}}  
\caption{
Scaling of the phase diffusion coefficient in the thermodynamic 
limit (\ref{eq:LT}), for a factor 5 variation of
the atom number $N$. 
Points with error bars: Simulation results.
The lines connect the points with the same $k_B T/\rho g$.
The points most on the left of the figure,
with $N=2.37\times 10^6$, are the ones of Fig.\ref{fig:varphi}, with a
grid size $32^3$. The other points are for a grid size $24^3$ ($N=10^6$)
and for a grid size $18^3$ ($N=4.22\times 10^5$).
\label{fig:finale_LT}}
\end{figure}

We now investigate the existence of a thermodynamical limit for the quantity $DN$, given that the
Bogoliubov limit is already reached. The thermodynamical limit is defined as usual as
\begin{equation}
N \to \infty \,, \:\:\: V \to \infty \, \:\:\:{\mbox{with}} \:\:\: \rho = \mbox{constant} \,,
\label{eq:LT}
\end{equation} 
the other parameters ($g,l_1,l_2,l_3,T$) being fixed. The result is shown in Fig.\ref{fig:finale_LT}
where $DN$ is constant within the error bars, over a factor 5 of variation of $N$ and 4 considered temperatures.

\begin{figure}[htb]
\centerline{\includegraphics[width=7cm,clip=]{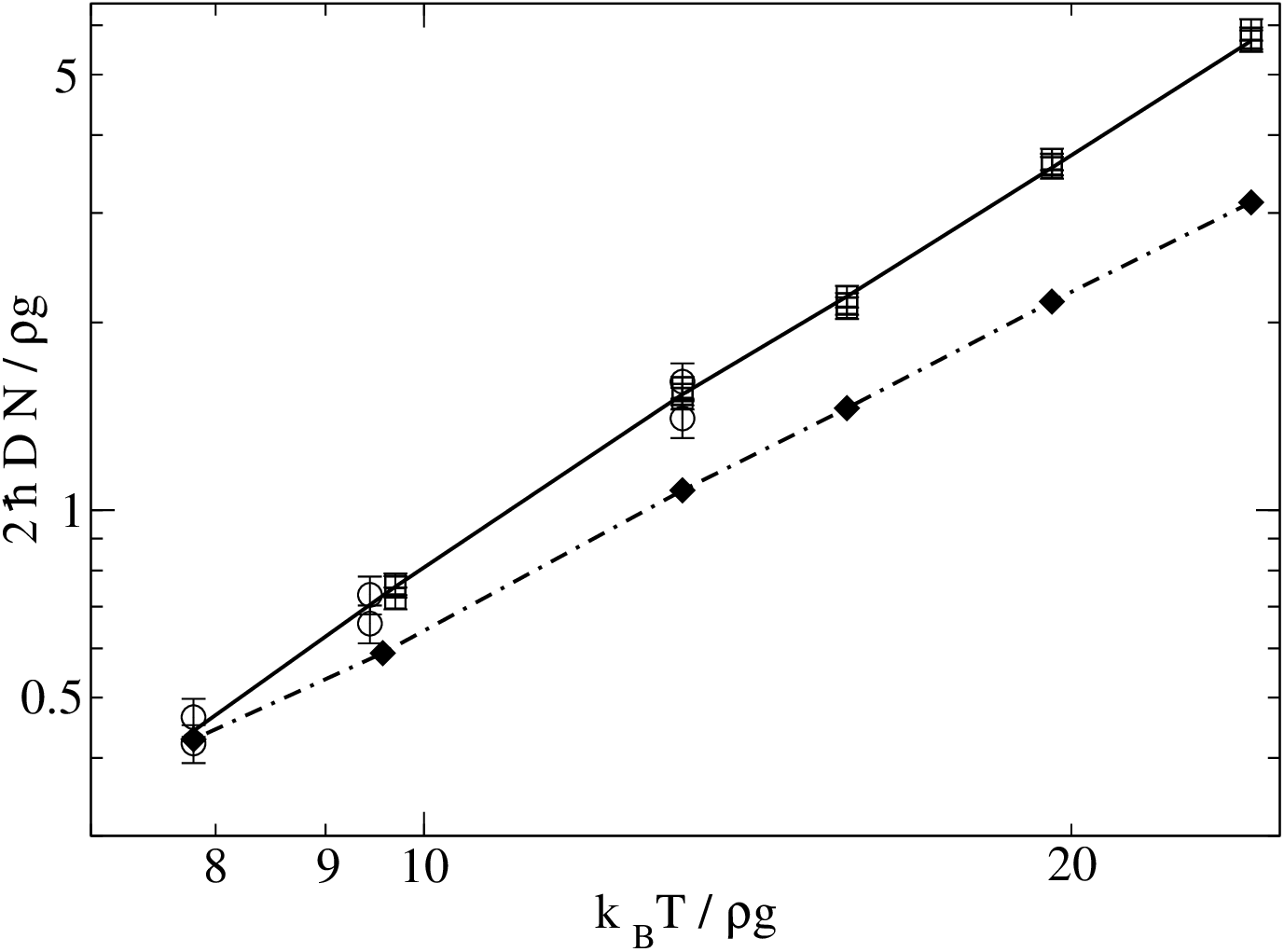}}  
\caption{Universal curve for the rescaled phase diffusion coefficient
as a function of $k_B T/\rho g$ in log-log scale.
Symbols with error bars: Simulation results.
Circles: Results of Fig.\ref{fig:finale_CD}. Squares:
Results of Fig.\ref{fig:finale_LT}.
The solid line connects the points with the largest value of $N$
of Fig.\ref{fig:finale_LT}, plus the average of the two points
of Fig.\ref{fig:finale_CD} with the lowest temperature.
Dashed-dotted line with filled diamonds: 
$D^{\rm approx}$ from the projected Gaussian approach. The $\Gamma_{\mathbf k}$ for the projected Gaussian approach are 
calculated on the same discrete grids used in the simulation points connected by the solid line.
\label{fig:bella}}
\end{figure}

In what follows, using dimensional analysis, we show that for our cut-off procedure
(\ref{eq:cutoff}), the dimensionless quantity $\hbar DN/\rho g$ is a function
of a single parameter $k_BT/\rho g$ once the Bogoliubov and thermodynamical limits are reached.
Six independent physical quantities are present in the model
\begin{equation}
\{ \, \hbar \,, \: m \,, \: g\, , \: V\,, \: k_B T\,, \: N \,\} \,.
\label{eq:indep}
\end{equation}
The lattice spacings $l_1,l_2,l_3$ are not independent parameters since their ratios are fixed
and their value is determined by (\ref{eq:cutoff}) once the quantities (\ref{eq:indep}) are fixed.
Equivalently we can replace $g$ by $\rho g$ and the volume $V$ by $k_B T_c$, where $T_c$ is the transition temperature of the ideal gas
given by
\begin{equation}
\rho \left( \frac{2\pi \hbar^2}{m k_B T_c}\right)^{3/2}=\zeta(3/2) \,.
\end{equation}
We then have
\begin{equation}
\frac{\hbar DN}{\rho g}=f\left(\hbar,m,\rho g,\frac{k_B T_c}{\rho g},\frac{k_B T}{\rho g}, N\right) \,.
\end{equation}
The three quantities $\hbar,m,\rho g$ can be recombined to form a length, a time and a mass which are three independent
dimensioned quantities. Since $f$ and its other three variables are dimensionless, $f$ does not depend of its first three
variables. In the thermodynamical limit $N \to \infty$ so the sixth variable of $f$ drops out of the problem. In the Bogoliubov
limit, $k_B T_c/\rho g \to \infty$ so that the forth variable of $f$ also drops.
We thus conclude that
\begin{equation}
\frac{\hbar DN}{\rho g}=f\left(\frac{k_B T}{\rho g}\right) \,.
\label{eq:Duniv}
\end{equation}
In Fig.\ref{fig:bella} we show the graph of $f$ as obtained by our classical field model
collecting all the simulation results of Fig.\ref{fig:finale_LT} and Fig.\ref{fig:finale_CD}. 
We have used a log-log scale in Fig.\ref{fig:bella} to reveal that the function $f$ is approximately
a power law in the considered range of $k_B T/\rho g$.

\section{Projected Gaussian approximation}
\label{sec:heuristic}
In this section we propose an approximate analytical formula for the phase diffusion coefficient that gives some physical insight
and can be extended to the quantum field case.

\subsection{Classical field}

We wish to calculate the integral of a correlation function 
${\cal{C}}(t)=\langle A(t) A(0) \rangle - \langle A(t) \rangle \langle A(0) \rangle$
of an observable $A$ of the form
\begin{equation}
A= \sum_{\mathbf{k}\neq\mathbf{0}} A_{\mathbf{k}} |b_{\bf k}|^2 \:,\:\:\:
\end{equation}
where $b_\mathbf{k}$ are the amplitudes of the field over the Bogoliubov modes.
We thus introduce the ${\cal M}\times{\cal M}$ covariance matrix $Q$ with matrix elements
\begin{equation}
Q_{\mathbf{k,k'}}(t)=\langle \delta n_{\mathbf k}(t) \delta n_{\mathbf k'}(0)   \rangle
\end{equation}
where $\delta n_{\mathbf k}= n_{\mathbf k} - \bar{n}_{\mathbf k}$ is the fluctuation of the occupation number of the
corresponding Bogoliubov mode and
${\cal M}=V/(l_1 l_2 l_3)-1$ is the number of Bogoliubov modes.
One thus has
\begin{equation}
{\cal C}(t)=\vec{A} \cdot Q(t) \vec{A}
\label{eq:Ccompact}
\end{equation}
where $\vec{A}$ is the vector of components $A_{\mathbf k}$. Since the system is described in this section by the microcanonical
ensemble for the Bogoliubov Hamiltonian, the matrix $Q$ obeys the relation 
\begin{equation}
\vec{\epsilon} \cdot Q = {}^t\vec{0}   \:\:\:\: \mbox{and} \:\:\:\: Q \vec{\epsilon} = \vec{0} \,,
\label{eq:contrainte}
\end{equation}
where the vector $\vec{\epsilon}$ collects the Bogoliubov energies.

By using the microcanonical classical averages \cite{hypersph}
one directly accesses the $t=0$ value of the matrix $Q$,
\begin{eqnarray}
\mbox{for} \, \:{\bf k}\neq{\bf k'} :&\:\:\:& Q_{\mathbf{k,k'}}(0) = 
 -\frac{\bar{n}_{\bf k}  \bar{n}_{\bf k'} }{{\cal M}+1} \\
\mbox{for} \, \:{\bf k}={\bf k'} : &\:\:\:& Q_{\mathbf{k,k}}(0)  = \bar{n}_{\bf k}^2 \; \frac{{\cal M}-1}{{\cal M}+1}\,.
\end{eqnarray}
Remarkably we can express this result in terms of the result one would have in the canonical ensemble
with average energy equal to the microcanonical energy,
adding a projector which suppresses energy fluctuations:
\begin{equation}
Q(0) = \frac{\cal M}{{\cal M}+1} \, P^\dagger Q^{\rm Gauss}(0) P  
\label{eq:proj}
\end{equation}
with
\begin{equation}
P_{\mathbf{k,k'}}=\delta_{\mathbf{k,k'}}-\epsilon_{\mathbf k} \alpha_{\mathbf{k'}} \,.
\label{eq:dyade}
\end{equation} 
The vector $\vec{\alpha}$ is adjoint to the vector $\vec{\epsilon}$ so that
$P \vec{\epsilon}=\vec{0}$. Its components are given by
\begin{equation}
\alpha_{\mathbf k}=\frac{1}{{\cal M} \epsilon_{\mathbf k}}
\end{equation}
and $Q^{\rm Gauss}$ is the value of the covariance matrix in the canonical ensemble 
\begin{equation}
Q^{\rm Gauss}_{\mathbf{k,k'}}(0)=\delta_{\mathbf{k,k'}} \bar{n}_{\mathbf{k}}^2 \,.
\end{equation}
The apex ``Gauss"
reminds the fact that the $b_{\mathbf k}$ have a Gaussian probability distribution in the canonical ensemble,
contrarily to the case of the microcanonical ensemble.

For the $t=0$ value of the phase derivative correlation function one then obtains
\begin{equation}
{\cal{C}}(0)= \frac{\cal M}{{\cal M}+1} \left[ \sum_{\mathbf{k}\neq\mathbf{0}} 
A_\mathbf{k}^2 \bar{n}_k^2 - \frac{1}{\cal M} \left( \sum_{\mathbf{k}\neq\mathbf{0}} A_\mathbf{k} \bar{n}_k \right)^2 \right]
\label{eq:C(0)}
\end{equation}
with
\begin{equation}
A_{\mathbf{k}}=-\frac{g}{V} (\tilde{U}_k+\tilde{V}_k)^2 \,.
\end{equation}
We verified that (\ref{eq:C(0)}), represented as a dashed line in Fig.\ref{fig:corr}, is in agreement with the numerical simulation as one enters the Bogoliubov limit (\ref{eq:CD}). 
Note that within the Bogoliubov aproximation
$b_{\bf k}(t) \simeq b_{\bf k}(0)\, e^{-i \tilde{\epsilon}_k t\hbar}$, the phase derivative correlation function remains equal to
its $t=0$ value (\ref{eq:C(0)}) at all times. This is in clear disagreement with the numerical simulation, and it would lead to
a ballistic spreading of the condensate phase.

Our approximate treatment consists in extending the relation (\ref{eq:proj}) at positive times, using the fact that in a Gaussian theory
one would have
\begin{equation}
Q^{\rm Gauss}_{\mathbf{k,k'}}(t)=Q^{\rm Gauss}_{\mathbf{k,k'}}(0) e^{-\Gamma_{\mathbf k} t} \,.
\label{eq:Q(t)Gauss}
\end{equation}
One indeed assumes in the Gaussian model
\begin{equation}
|\langle b_{\bf k}^\ast(t)  b_{\bf k'}(0) \rangle_{\rm Gauss}| = \delta_{\mathbf{k,k'}} \bar{n}_k e^{-\Gamma_{\mathbf k} t/2} 
\label{eq:corrbk}
\end{equation}
and one uses Wick theorem to obtain (\ref{eq:Q(t)Gauss}).
Physically equation (\ref{eq:corrbk}) describes Landau Beliaev processes that decorrelate the $b_{\bf k}$. 
It can be derived for example with a master equation approach as done in an appendix of \cite{PRAEmilia}.

For the phase derivative correlation function one then obtains the approximate expression
\begin{eqnarray}
{\cal{C}}^{\rm approx}(t)&=& \frac{\cal M}{{\cal M}+1} \left[ \sum_{\mathbf{k}\neq\mathbf{0}} 
A_\mathbf{k}^2 \bar{n}_k^2  e^{-\Gamma_{\mathbf k} t} \right. \nonumber \\
&-& \left.  \frac{2}{\cal M} \left( \sum_{\mathbf{k}\neq\mathbf{0}} A_\mathbf{k} \bar{n}_k \right) 
\left( \sum_{\mathbf{k'}\neq\mathbf{0}} A_\mathbf{k'} \bar{n}_{k'} e^{-\Gamma_{\mathbf{k'}} t} \right)  \right.  \nonumber\\
&+& \left. 
 \frac{1}{{\cal M}^2}  \left( \sum_{\mathbf{q}\neq\mathbf{0}} e^{-\Gamma_{q} t} \right)  \left( \sum_{\mathbf{k}\neq\mathbf{0}} A_\mathbf{k} \bar{n}_k \right)^2 \right] \,.
\label{eq:C(t)}
\end{eqnarray}
We represent (\ref{eq:C(t)}) as a dashed-dotted line in Fig.\ref{fig:corr}. The resulting approximation on the diffusion coefficient
is obtained by integration
\begin{equation}
D^{\rm approx}=\int_0^{+\infty} \, {\cal C}^{\rm approx}(t) \, dt \,.
\label{eq:Dapprox}
\end{equation}
In Fig.\ref{fig:bella} we compare the approximation (\ref{eq:Dapprox}) (diamonds linked by a dashed-dotted line) to the numerical simulation results.
The agreement is acceptable in the considered range of $k_BT/\rho g$. The Landau-Beliaev damping rates $\Gamma_{\mathbf k}$ are calculated on
the same discrete grid as the simulation points as explained in \cite{Cartago}.

\subsection{Quantum field}
In this subsection we extend the approximate formula for the phase diffusion coefficient to the quantum case.
The Bogoliubov amplitudes and occupation numbers are now operators $\hat{b}_{\mathbf k}$, $\hat{n}_{\mathbf k}$.
As in the classical case we introduce the covariance matrix of the Bogoliubov occupation numbers
\begin{equation}
Q_{\mathbf{k,k'}}(t)=\langle \delta \hat{n}_{\mathbf k}(t) \delta \hat{n}_{\mathbf k'}(0)   \rangle \,.
\end{equation}
To obtain the $t=0$ value of $Q$ we need to compute quantum averages in the microcanonical ensemble. To this end
we use a result derived in \cite{PRAEmilia} giving the first deviation between the microcanonical expectation value $\langle{O}\rangle$
and the canonical one $\langle O\rangle_{\rm can}(T)$ in the limit of a large system for an arbitrary observable ${O}$: 
\begin{equation}
\langle{O}\rangle - \langle O\rangle_{\rm can}(T) =
-\frac{1}{2} k_B T^2 \frac{d}{dT}\left(\frac{d\langle O\rangle_{\rm can}/dT}{d\langle H\rangle_{\rm can}/dT}\right)
+\ldots
\end{equation}
where the canonical temperature $T$ is such that the mean energy $\langle H\rangle_{\rm can}$ in the canonical ensemble is equal to the microcanonical energy $E$.

For ${\mathbf k \neq \mathbf {k'} }$, using $d\bar{n}_k/dT=\epsilon_k \bar{n}_k (\bar{n}_k+1)/k_B T^2$, one finds
\begin{equation}
Q_{\mathbf{k,k'}}(0) \simeq - \frac{\epsilon_k \epsilon_{k'} \bar{n}_k(\bar{n}_k+1) \bar{n}_{k'}(\bar{n}_{k'}+1)}
{\sum_{\mathbf q\neq \mathbf{0}} \epsilon_q^2 \bar{n}_q (\bar{n}_q+1)}.
\end{equation}
This scales as $1/\mathcal{M}$ in the thermodynamic limit. Since the number of off-diagonal terms of $Q$
in (\ref{eq:Ccompact}) is about $\mathcal{M}$ times larger than the number of diagonal terms of $Q$,
we have for consistency to calculate the diagnal terms of $Q$ up to order $1/\mathcal{M}^0$, that is
the deviation from the canonical value $\bar{n}_k (\bar{n}_k+1)$ is not required.
To exactly obtain the energy conservation (\ref{eq:contrainte}), it is however convenient to include, rather than the exact
deviation between the canonical and microcanonical values,
an {\sl ad hoc} approximate correction of order $1/\mathcal{M}$:
\begin{equation}
Q_{\mathbf{k,k}} \simeq \bar{n}_k (\bar{n}_k+1) - \frac{\epsilon_k^2 \bar{n}_k^2 (\bar{n}_k+1)^2}
{\sum_{\mathbf q\neq \mathbf{0}} \epsilon_q^2 \bar{n}_q (\bar{n}_q+1)}
\end{equation}
In this way, we recover the structure of (\ref{eq:proj}), where the $t=0$ value of $Q$ is deduced from
the one in the canonical ensemble,
\begin{equation}
Q^{\rm Gauss}_{\mathbf{k,k'}}(t=0)= \bar{n}_k (\bar{n}_k+1) \delta_{\mathbf{k,k'}}
\end{equation}
by the action of a projector $P$,
\begin{equation}
Q(t=0) \simeq P^\dagger Q^{\rm Gauss}(t=0) P.
\label{eq:init_q}
\end{equation}
The projector $P$ still involves the dyadic structure (\ref{eq:dyade}), with a new expression
for the vector $\vec{\alpha}$:
\begin{equation}
\alpha_\mathbf{k} = \frac{\epsilon_k \bar{n}_k (\bar{n}_k+1)}{\sum_{\mathbf q\neq \mathbf{0}} \epsilon_q^2 \bar{n}_q (\bar{n}_q+1)}.
\end{equation}
As a check, one can apply the classical field limit to the above  quantum expressions.
One recovers (\ref{eq:proj}), apart from the global factor
$\mathcal{M}/(\mathcal{M}+1)$, whose deviation from unity
gives rise to terms beyond the accuracy of the present calculation.

At positive times, our quantum projected Gaussian approximation assumes that (\ref{eq:init_q})
still holds, 
\begin{equation}
Q^{\rm approx}(t) \simeq P^\dagger Q^{\rm Gauss}(t) P
\end{equation}
with the Gaussian covariance matrix
\begin{equation}
Q^{\rm Gauss}_{\mathbf{k,k'}}(t) = \delta_{\mathbf{k,k'}}  \bar{n}_k (\bar{n}_k+1) e^{-\Gamma_{\mathbf k} t}
\end{equation}
where the Landau-Beliaev damping rate $\Gamma_{\mathbf k}$ is now the usual one, that is for the quantum field theory.
From (\ref{eq:Ccompact}) one obtains an approximate expression for the phase derivative correlation
function, and from (\ref{eq:Dapprox}) an approximate expression for the quantum field phase diffusion
coefficient \cite{symmetric}:
\begin{equation}
D^{\rm approx} = \sum_{k\neq\mathbf{0}} [(P \vec{A}\,)_{\mathbf{k}}]^2 \frac{\bar{n}_k(\bar{n}_k+1)}{\Gamma_{\mathbf k}}
\label{eq:qDapprox}
\end{equation}
where the projection of the vector $\vec{A}$ was introduced:
\begin{multline}
(P \vec{A}\,)_{\mathbf{k}}=   -\frac{g}{V}\times \\ \left[ (U_k+V_k)^2 
-  \epsilon_k 
\frac{\sum_{\mathbf q\neq \mathbf{0}} \epsilon_q (U_q+V_q)^2\bar{n}_q (\bar{n}_q+1)}
{\sum_{\mathbf q\neq \mathbf{0}} \epsilon_q^2 \bar{n}_q (\bar{n}_q+1)}\right].
\end{multline}
In this expression, the modes amplitudes $U_k,V_k$ and energy $\epsilon_k$ have the usual expressions
of the quantum field Bogoliubov theory,
\begin{eqnarray}
U_k+V_k &=& \frac{1}{U_k-V_k} = \left(\frac{\hbar^2 k^2/2m}{2\rho g + \hbar^2 k^2/2m}\right)^{1/4} \\
\epsilon_k &=& \left[\frac{\hbar^2k^2}{2m}\left(2\rho g+\frac{\hbar^2k^2}{2m}\right)\right]^{1/2}.
\end{eqnarray}
In appendix \ref{app:gammas} we give the explicit expression of the Landau-Beliaev damping rates $\Gamma_k$
and the approximate phase diffusion coefficient in the thermodynamical limit. The existence of such a limit
is due to the fact that none of the momentum integrals involved are infrared divergent, keeping in mind
that $(U_k+V_k)^2$, $\epsilon_k$ and $\Gamma_k$ vanish linearly with $k$
\cite{Giorgini}, while $\bar{n}_k(\bar{n}_k+1)$ diverges as $1/k^2$.
As expected from the analysis of the previous section, the scaled diffusion coefficient $\hbar D^{\rm approx} N/\rho g$ 
is a function of $k_B T/\rho g$ only, see (\ref{eq:Djolie}).

\begin{figure}[htb]
\centerline{\includegraphics[width=7cm,clip=]{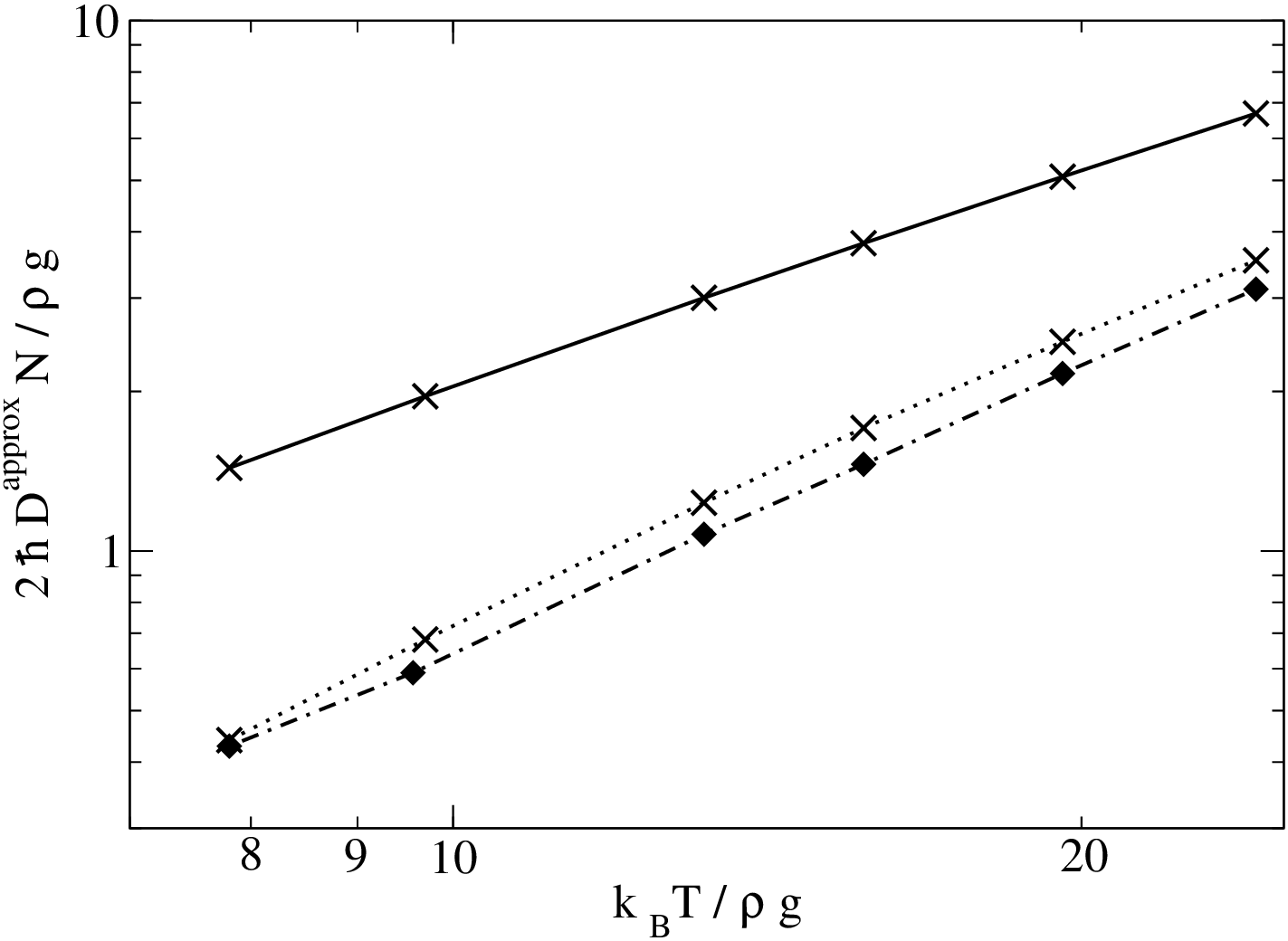}}  
\caption{Approximate phase diffusion coefficient in the thermodynamic limit
as a function of $k_B T/\rho g$ in log-log scale.
Solid line with crosses: $D^{\rm approx}$ for the quantum field from Eq.(\ref{eq:Djolie}).
Dotted line with crosses: Effect of an energy cut-off equal to $k_B T$ in Eq.(\ref{eq:Djolie}).
Dashed-dotted line with filled diamonds: $D^{\rm approx}$ for the classical field from Eq.(\ref{eq:Dapprox}) (same data as
in Fig.\ref{fig:bella}). 
\label{fig:fig_quant}}
\end{figure}
In Fig.\ref{fig:fig_quant} we show the values of $D^{\rm approx}$ in the thermodynamic limit for the same values of 
$k_BT/\rho g$ as in Fig.\ref{fig:bella}. To see the effect of an energy truncation at $k_B T$,
we also show the values of $D^{\rm approx}$ obtained
by introducing an energy cut-off $\epsilon_q < k_B T$ in (\ref{eq:Djolie}) and the same cut-off $\epsilon_k<k_B T$
in the integrals (\ref{eq:GL},\ref{eq:GB}) giving the damping rate $\Gamma_q$.
As expected,  the resulting values of $D^{\rm approx}$ are close to the values of $D^{\rm approx}$ obtained
for the classical field model in Fig.\ref{fig:bella}, and that we have reported in Fig.\ref{fig:fig_quant} for comparison. We conclude
that the diffusion coefficient is indeed affected by an energy cut-off, and the coefficient obtained in the classical field simulations with an energy cut-off $k_BT$ might differ quantitatively from the real one by a factor of about two.

\section{Conclusion}
\label{sec:conclusion}
Using a classical field model, we have shown that the phase of a Bose-Einstein condensate undergoes true diffusion in time,
when the gas is initially prepared in the microcanonical ensemble.
Parametrizing the microcanonical energy $E$ by the
temperature $T$ of the canonical ensemble with average energy $E$, we could show that the rescaled diffusion coefficient $\hbar D N/\rho g$, where $N$ is the fixed number of particles and $\rho g$ is the Gross-Pitaevskii chemical potential,
is a function of a single variable $k_B T/\rho g$ in the double thermodynamic and Bogoliubov limit.

We have derived an approximate formula for the diffusion coefficient,
in fair agreement with the classical field simulations. 
We could generalize the approximate formula to the quantum field case, show that it also admits a thermodynamic limit and 
that it satisfies the scaling property found for the classical field. 
We have used the quantum approximate formula to evaluate 
the effect of an energy cut-off, not required in the quantum theory and unavoidable in the classical field model.

The perspective of using the condensate phase spreading
to experimentally distinguish among different statistical ensembles is fascinating,  although the measurement of
the intrinsic phase diffusion of a Bose-Einstein condensate discussed here remains a challenge and will be the
subject of further investigations.

\section{Acknowledgments} The authors would like to acknowledge E. Witkowska for her initial contribution to the project,
K. M\o lmer for hospitality and useful discussions and F. Hulin-Hubard for valuable help with computers. A.S. acknowledges stimulating discussions with W. Phillips, K. Rz\c{a}\.zewski, M. Gajda and M. Oberthaler. Our groups are members of IFRAF.

\appendix

\section{Landau-Beliaev damping rates and approximate phase diffusion coefficient in the thermodynamic limit}
\label{app:gammas}
We start with the Landau damping rate of a Bogoliubov mode of wave vector $\mathbf{q}$ as given in \cite{PRAEmilia}
\begin{equation}
\Gamma_q^L=\frac{g^2\rho}{\pi^2\hbar} \int d^3 k \,
L_{k,k'}^2 (\bar{n}_k-\bar{n}_{k'})
\delta(\epsilon_q+\epsilon_k-\epsilon_{k'})
\label{eq:GL}
\end{equation}
with
\begin{equation}
L_{k,k'} = U_q V_{k} U_{k'} + (U_q+V_q)(U_{k} U_{k'} +V_{k} V_{k'}) + V_q U_{k} V_{k'}.
\end{equation}
The mode of wave vector $\mathbf{q}$ scatters an excitation of wave vector $\mathbf{k}$ giving rise to an excitation
of wave vector $\mathbf{k}'$.
The final mode has to satisfy momentum conservation so that $\mathbf{k}'=\mathbf{k}+\mathbf{q}$. Energy conservation 
$\epsilon_q+\epsilon_{k}=\epsilon_{k'}$ is ensured by the delta distribution in (\ref{eq:GL}). In the integral over $\mathbf{k}$ we use
spherical coordinates of axis $\mathbf{q}$, $\theta$ being the polar angle.
We introduce the momentum $\check{q}$ scaled by the inverse of  the healing length $\xi$ and the mode energy $\check{\epsilon}_q$
scaled by the Gross-Pitaevskii chemical potential $\rho g$:
\begin{eqnarray}
 \check{q}&=& q \left(\frac{\hbar^2}{2m\rho g}\right)^{1/2} = q \xi,\\
\check{\epsilon}_{q} &=&\frac{\epsilon_q}{\rho g}  = [\check{q}^2(\check{q}^2+2)]^{1/2} \,.
\end{eqnarray}
As a consequence, the mean occupation number $\bar{n}_q$ is a function of $\check{q}$ and of the ratio
$k_BT/\rho g$ only, and the mode amplitudes $U_q, V_q$ are functions of $\check{q}$ only. 
Introducing the notation $u= \cos \theta$, one has
 \begin{equation}
 \delta(\check{\epsilon}_q+\check{\epsilon}_k-\check{\epsilon}_{k'}) = \delta(u-u_0^L)\frac{\check{\epsilon}_{q}+\check{\epsilon}_{k}}
 {2\check{k}\check{q}[1+(\check{\epsilon}_{q}+\check{\epsilon}_{k})^2]^{1/2}}
 \end{equation}
 where 
 \begin{equation}
 u_0^L = \frac{\left[ 1+(\check{\epsilon}_q+\check{\epsilon}_k)^2\right]^{1/2}
 -\left( 1+\check{q}^2+\check{k}^2\right)}{2\check{k}\check{q}}.
 \end{equation}
 One can show that $u_0^L$ is in between $-1$ and $1$ for all values of $\check{k}$
 and $\check{q}$, so that the angular integration is straightforward and leads to
\begin{equation}
{\Gamma}_q^L=  \frac{g}{\pi \hbar\xi^3}
\int_0^{+\infty} \!\! d\check{k} \,  \, L_{k,k'}^2 
 \frac{\check{k}(\check{\epsilon}_k+\check{\epsilon}_q) (\bar{n}_k-\bar{n}_{k'})}
{\check{q}\left[ 1+(\check{\epsilon}_q+\check{\epsilon}_k)^2\right]^{1/2}}
\end{equation}
with 
\begin{equation}
1+\check{k}^{'2}=\left[ 1+(\check{\epsilon}_q+\check{\epsilon}_k)^2\right]^{1/2}.
\end{equation}

A similar procedure may be applied to the Beliaev damping rate for the mode $\mathbf{q}$.
From \cite{PRAEmilia} one has
\begin{equation}
\Gamma_q^B=\frac{g^2\rho}{2\pi^2\hbar} \int d^3 k
B_{k,k'}^2 (1+\bar{n}_k+\bar{n}_{k'}) 
\delta(\epsilon_k+\epsilon_{k'}-\epsilon_q)
\label{eq:GB}
\end{equation}
with
\begin{equation}
B_{k,k'} = U_q U_{k} U_{k'} + (U_q+V_q)(V_{k} U_{k'} +U_{k} V_{k'}) + V_q V_{k} V_{k'}.
\end{equation}
Here the mode of wave vector $\mathbf{q}$ decays into an excitation of wave vector $\mathbf{k}$
and an excitation of wave vector $\mathbf{k}'$. Momentum conservation imposes
$\mathbf{k}'=\mathbf{q}-\mathbf{k}$. Energy conservation $\epsilon_{k'}=\epsilon_q-\epsilon_k$
is ensured by the delta distribution in (\ref{eq:GB}), and clearly imposes $k < q$.
With the same scaled variables and spherical coordinates as above, one obtains
\begin{equation}
\delta(\check{\epsilon}_k+\check{\epsilon}_{k'}-\check{\epsilon}_q) =
\delta(u-u_0^B)\frac{\check{\epsilon}_q-\check{\epsilon}_k}{2\check{k}\check{q}
[1+(\check{\epsilon}_q-\check{\epsilon}_k)^2]^{1/2}}
\end{equation}
where
\begin{equation}
u_0^B = \frac{1+\check{q}^2+\check{k}^2-[1+(\check{\epsilon}_q-\check{\epsilon}_k)^2]^{1/2}}{2\check{k}\check{q}}.
\end{equation}
One can show that $u_0^B$ is in between $-1$ and $1$, whatever the values of $\check{q}$
and $\check{k}<\check{q}$, so that angular integration is straightforward and gives
\begin{equation}
\Gamma_q^B = \frac{g}{2\pi\hbar\xi^3} \int_0^{\check{q}} d\check{k}\,B_{k,k'}^2 
\frac{\check{k}(\check{\epsilon}_q-\check{\epsilon}_k)(1+\bar{n}_k+\bar{n}_{k'})}{\check{q}[1+(\check{\epsilon}_q-\check{\epsilon}_k)^2]^{1/2}}
\end{equation}
with
\begin{equation}
1+\check{k}^{'2} = \left[1+\left(\check{\epsilon}_q-\check{\epsilon}_k\right)^2\right]^{1/2}.
\end{equation}
Finally we introduce the rescaled total damping rate,
\begin{equation}
\check{\Gamma}_q = \frac{2\pi^2\hbar\xi^3}{g} \left(\Gamma_q^L + \Gamma_q^B\right),
\end{equation}
a dimensionless function of $k_B T/\rho g$ only.
From (\ref{eq:qDapprox}) one then obtains in the thermodynamic limit an approximate expression for the phase
diffusion coefficient depending only on $k_B T/\rho g$,
\begin{equation}
\frac{\hbar D^{\rm approx} N}{\rho g}=\int_0^{+\infty} d\check{q} \, \check{q}^2 \, ({\cal A}_{q}^P)^2 \frac{\bar{n}_q(1+\bar{n}_q)}
{\check{\Gamma}_q}
\label{eq:Djolie}
\end{equation}
with
\begin{equation}
{\cal A}_{q}^P=(U_q+V_q)^2 - \check{\epsilon}_q 
\frac{\int_0^{+\infty} d\check{k} \, \check{k}^2 \, \check{\epsilon}_k (U_k+V_k)^2 \bar{n}_k (\bar{n}_k+1)}
{\int_0^{+\infty} d\check{k} \, \check{k}^2 \, \check{\epsilon}_k^2 \bar{n}_k (\bar{n}_k+1)} \,.
\end{equation}


\begin{thebibliography}{99}
\bibitem{Delta} $\Delta$ stands here for the operator acting on functions
on the lattice, such that its eigenvectors are the plane waves $e^{i\mathbf{k}\cdot
\mathbf{r}}$ with eigenvalues $-k^2$.
\bibitem{EBvarphi}  We calculate $\mathrm{Var} \, \varphi(t)$ and its error bars for 5000 time points. 
These 5000 time points are collected into 100 packets of 50 consecutive points, the $i^{th}$ packet corresponding to a time interval $[t_i,t_i+\Delta t]$. $\Delta t$ is chosen of the order of the typical correlation time of
the condensate phase derivative (see Fig.\ref{fig:corr}). 
A temporal average of ${\mathrm{Var} \, \varphi}$ over each time interval is then calculated. and reported in the figure as a single point at $t_i+\Delta t/2$ with its mean square (time averaged) error bar.
\bibitem{osc_terms} As explained in \cite{PRAEmilia}, in (\ref{eq:phidotnonosc}) we assume a small non condensed
fraction and we neglect an oscillatory part. 
We checked analytically, within the Bogoliubov approach and for an infinite cut-off, that the contribution of the oscillatory part of $\dot{\varphi}$
to the correlation function $\mathcal{C}(t)$ is a rapidly decreasing function over the time scale $\hbar/\rho g$.
\bibitem{bk}
In our classical field model, the field amplitudes on the Bogoliubov modes are given by
$b_{\mathbf{k}}(t) = dV\sum_\mathbf{r} 
\tilde{U}_k \frac{e^{-i\mathbf{k}\cdot\mathbf{r}}}{\sqrt{V}} e^{-i\theta(t)} \psi_\perp(\mathbf{r},t) 
+
\tilde{V}_k \frac{e^{i\mathbf{k}\cdot\mathbf{r}}}{\sqrt{V}} e^{i\theta(t)} \psi_\perp^*(\mathbf{r},t) 
$
where $\psi_\perp$ is the component of $\psi$ orthogonal to the condensate mode and
the real amplitudes $\tilde{U}_k$, $\tilde{V}_k$, normalized
as $\tilde{U}_k^2-\tilde{V}_k^2=1$, are given by the usual Bogoliubov theory, here with 
the modified dispersion relation
$\tilde{U}_k+\tilde{V}_k = [{\tilde{E}_k}/({\tilde{E}_k+2\rho g})]^{1/4}$.
\bibitem{proviso} Taking this limit in its true mathematical sense, we eventually reach
a {\sl discrete} regime where the Bogoliubov eigenstates do not form a quasi-continuum,
since the energy spacings between the Bogoliubov energy levels are fixed in the Bogoliubov
limit whereas the coupling amplitudes among them vanish in this limit. The concept of Beliaev-Landau
damping rates may be not appropriate in this discrete regime. To be more rigorous mathematically
one should first take the thermodynamic limit, and then take the Bogoliubov limit (now formulated
as $\rho\to+\infty$ for a fixed $\rho g$). Here we have kept the opposite order for pedagogical reasons,
the intuitive reasoning in the Bogoliubov limit allowing to infer the proportionality of $D$ with $1/N$.
\bibitem{symmetric} One may note  that our approximation for $Q(t)$ is a symmetric matrix,
which is not necessarily true for the exact $Q$ at $t>0$.
\end{thebibliography}
\end{document}